\documentclass{optica-article}

\journal{opticajournal}

\articletype{Research Article}

\usepackage{lineno}

\begin{document}

\title{Optical Frequency Averaging of Light}

\author{William Loh,\authormark{1,*} Ryan T. Maxson,\authormark{1} Alexander P. Medeiros,\authormark{1} Gavin N. West,\authormark{2} Paul W. Juodawlkis,\authormark{1} and Robert P. McConnell\authormark{1}}

\address{\authormark{1} Lincoln Laboratory, Massachusetts Institute of Technology, 244 Wood Street, Lexington, MA 02421, USA}
\address{\authormark{2} Massachusetts Institute of Technology, 77 Massachusetts Avenue, Cambridge, MA 02139, USA}

\email{\authormark{*} William.loh@ll.mit.edu} 



\begin{abstract}
The use of averaging has long been known to reduce noise in statistically independent systems that exhibit similar levels of stochastic fluctuation. This concept of averaging is general and applies to a wide variety of physical and man-made phenomena such as particle motion, shot noise, atomic clock stability, measurement uncertainty reduction, and methods of signal processing. Despite its prevalence in use for reducing statistical uncertainty, such averaging techniques so far remain comparatively undeveloped for application to light. We demonstrate here a method for averaging the frequency uncertainty of identical laser systems as a means to narrow the spectral linewidth of the resulting radiation. We experimentally achieve a reduction of frequency fluctuations from 40 Hz to 28 Hz by averaging two separate laser systems each locked to a fiber resonator. Critically, only a single seed laser is necessary as acousto-optic modulation is used to enable independent control of the second path. This technique of frequency averaging provides an effective solution to overcome the linewidth constraints of a single laser alone, particularly when limited by fundamental noise sources such as thermal noise, irrespective of the spectral shape of noise.
\end{abstract}

\section{Introduction}
The average of N identical fluctuating variables results in an ensemble whose mean fluctuations are reduced by \begin{math} \sqrt{N} \end{math}. This reduction in noise makes techniques of averaging inherently powerful for improving the signal-to-noise of datasets, for capturing images in a noisy background, and for measuring weak signals buried within noise. Ideally, such techniques of averaging would also be made applicable to laser frequency as the statistical fluctuations of frequency are of paramount importance to a host of applications including optical atomic clocks \cite{Hinkley2013, Bloom2014, Godun2014, Huntemann2016, Koller2017, Brewer2019}, gravitational wave detection \cite {Abramovici1992, Abbott2016}, quantum computing \cite {Cirac1995}, precision spectroscopy \cite {Rafac2000}, long-range lidar, and ultralow-noise optical frequency division \cite {Fortier2011}. The need to prepare multiple identical systems for averaging presents a trade-off between physical volume and laser performance. Yet, for many applications in basic science, laser linewidth is the primary metric and space concerns are nearly inconsequential. There, averaging of multiple laser systems offers one potential path towards reducing linewidths further, beyond the performance achievable by a single cavity alone. For other applications, compactness may be a more substantial concern, and the use of multiple systems for averaging can become somewhat costly in terms of size. However, in certain situations, such as for portable optical-atomic clocks, chip-integrated lasers are on the verge of meeting the linewidths necessary to address individual atoms \cite{Liu2022}. These lasers have already reached the limit of total available chip space, and thus averaging of multiple chips may enable integrated laser systems to cross the threshold for use in atomic physics.

While it is relatively trivial to apply averaging to quantities such as voltage or power, the real-time averaging of optical frequency is far more complex. This operation requires the summation of an array of optical frequencies, which would normally necessitate the use of nonlinear optics. As a consequence, only primitive techniques providing an effective form of averaging have been developed for optical systems, with greatly limited capabilities and scope. Such techniques rely on expanding the optical mode volume to provide an equivalent spatial averaging of thermal noise, specifically for application to optical resonators \cite {Gorodetsky2004, Notcutt2006, Matsko2007, Huang2019, Panuski2020}. Here, we instead demonstrate a universal method of averaging, applicable to any optical signal, by directly performing operations on the optical signals themselves. We circumvent the requirement of nonlinear optics by first converting optical frequency to voltage and subsequently performing the operations in voltage before converting back to an optical frequency.


\begin{figure}[t b !]
\centering
\includegraphics[width = 0.8 \columnwidth]{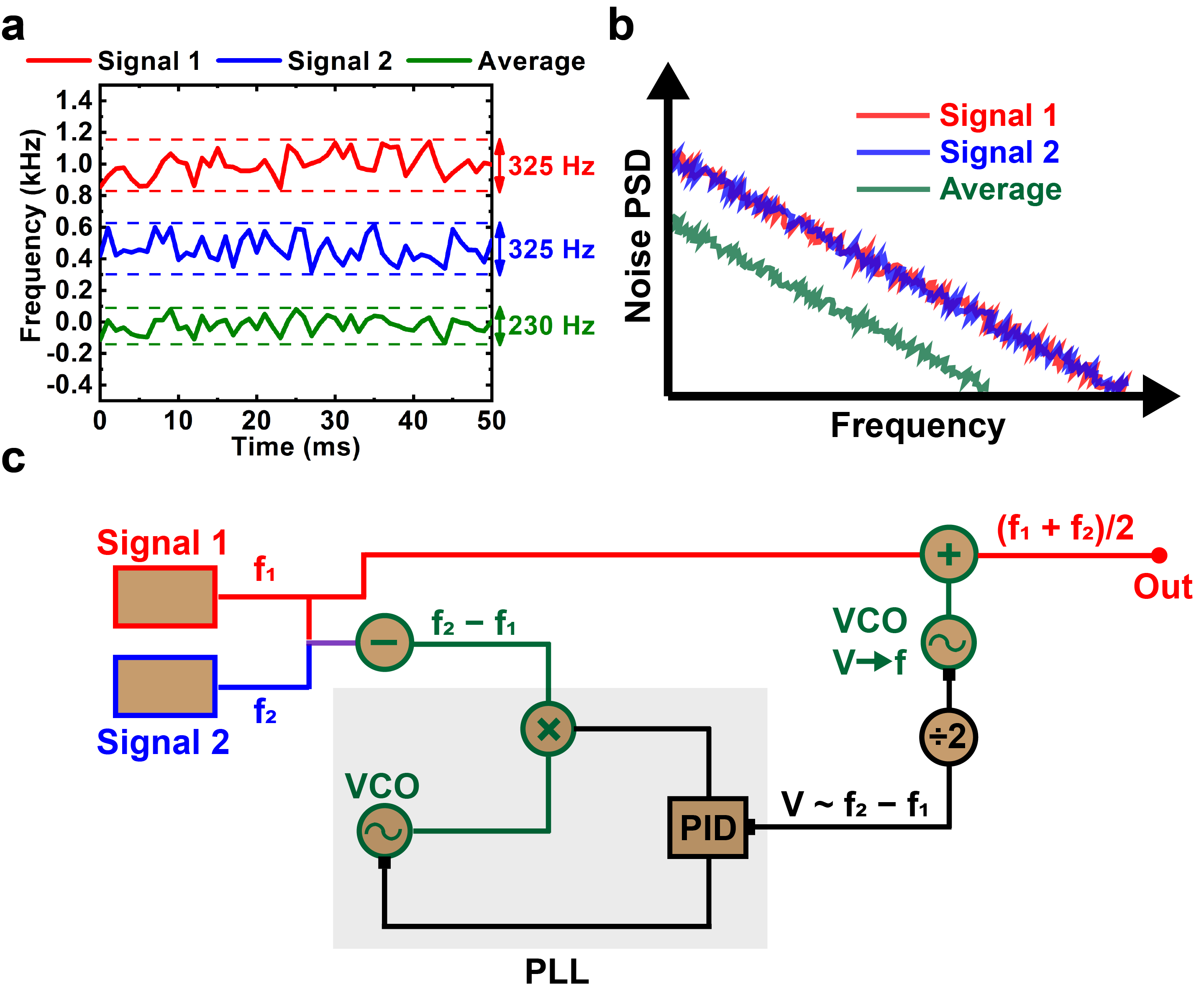}
\caption{
    \textbf{Optical Frequency Averaging Concept.}
    \textbf{a}, Experimental time series taken of two separate optical signals (red and blue) and their calculated average (green). The fluctuations are reduced by \begin{math} \sqrt{2} \end{math} in the average.
    \textbf{b}, Illustration of the power spectral density (PSD) of noise for two optical signals and the reduction gained through averaging. 
    \textbf{c}, Block diagram of the general system used for achieving averaging of two optical signals. The optical frequencies are first converted to a voltage through a phase-locked loop (PLL) where mathematical operations are easy to perform. The processed voltage is converted back to a frequency through a voltage controlled oscillator (VCO).
    \textbf{d}, Illustration of the optical signals used in averaging. These signals are general and can be pre-stabilized lasers that are locked to reference cavities.
}
\label{fig:fig1}
\end{figure}

With two identical input lasers, the resulting average exhibits frequency excursions that are \begin{math} \sqrt{2} \end{math} lower than that of each laser individually. Figure 1a shows this effect using experimental time traces collected from two separate laser sources. The details of these lasers will be discussed in Section 2. Each laser exhibits excursions on the scale of 325 Hz over 50 seconds. The numerically calculated average of their frequencies is a time trace exhibiting frequency excursions of 230 Hz, which is a factor of \begin{math} \sqrt{2} \end{math} lower. Figure 1b provides an illustrative example of the reduction in noise power spectral density before and after averaging. The averaging process decreases the noise uniformly across frequency, even for spectral shapes that are colored. However, in contrast to Fig. 1a, the reduction in power spectral density is a full factor of 2, due to the squaring of the noise processes inherent in the calculation of PSD.

Central to our implementation of optical frequency averaging is the conversion of optical frequency to a voltage, where mathematical operations are significantly more feasible. Figure 1c shows a block diagram of the general steps involved in this process. The heterodyne of the two individual optical signals to be averaged is first used to generate a microwave signal beat note. A low-noise voltage controlled oscillator is phase-locked to this microwave frequency to imprint a replica of the frequency difference between the two optical sources onto the control voltage. The voltage is next halved and sent back to a second VCO with identical frequency-tuning characteristics. The resulting microwave frequency \begin{math} (f_2-f_1)/2 \end{math} is summed with the original laser frequency \begin{math} f_1 \end{math} to achieve a mathematical averaging of the original two optical signals \begin{math} (f_1+f_2)/2 \end{math}. Alternatively, a microwave frequency divider can be used instead of a phase-locked loop, which accomplishes the division by 2 directly in the microwave frequency. Note that optical signals 1 and 2 participating in the averaging are general and can take on the form of lasers that are already stabilized to reference cavities \cite {Young1999, Jiang2011, Kessler2012, Lee2013, Loh2015, Loh2020, Zhang2020}. This is powerful as it enables one to first achieve a significant linewidth reduction through pre-stabilization techniques before the averaging process is performed to further narrow the linewidth. 

\section{Results and Measurement}

Figure 2a shows the circuit diagram used for implementing the optical frequency averaging system. A single seed laser is split into two paths, with each path to be stabilized to a separate ultralow-noise fiber cavity \cite {Loh2019}. The fiber cavity is made from splicing the ends together of an ultralow-loss evanescent coupler with a fixed coupling ratio of 5\% and a total fiber length of 2 meters. These stabilized lasers represent the two optical signals that will be averaged together to form the final system output. The primary path is further split two ways, with 50\% of the light passing through an AOM that performs the computation of averaging. The other 50\% of the signal passes through an EOM for generating phase modulation sidebands and subsequently through an SOA for amplification. The amplified signal is sent to one of the fiber resonators, and its output is photodetected and demodulated to create the error signal for a Pound-Drever-Hall (PDH) servo \cite {Drever1983}. This servo actuates on the laser current to keep the central laser frequency on resonance. This stabilized laser then serves as the seed laser for the second pathway which passes through another AOM to enable independent control of the light. This output is sent through the combination of an EOM and SOA before entering into a second ultralow-noise fiber resonator. The output is photodetected and demodulated, with the servo actuating on the AOM drive frequency. 

The circuit of Fig. 2a accomplishes two purposes. First, it creates two ultralow-noise lasers whose frequency fluctuations are each independently governed by the noise of their respective stabilization cavities. Critically, these cavities are limited by thermorefractive noise, which is statistically independent between the two cavities, and therefore as we later show, can be averaged down. Second, because the laser stabilized to the first cavity is used as the seed for the second cavity, the control voltage sent to the AOM takes on the differential frequency fluctuations between the two independent laser paths \begin{math} (f_2-f_1) \end{math}. This control voltage is directly sent into a second VCO that has half the frequency response. This turns the voltage back into an RF frequency and also accomplishes the division by 2. The resulting RF signal is directly sent to an AOM for combining back with the original optical signal to perform the optical frequency averaging operation. The frequency and voltage PSDs highlight the one-to-one correspondence in their spectral shapes, which enables operations to be performed in voltage before the signal is converted back to frequency. Figure 2b shows the sweeps over the two individual resonators along with their associated PDH error signals. The resonators each exhibit a linewidth of 1.1 MHz and are separated by 200 MHz, which corresponds to the frequency shift of the AOM. 


\begin{figure}[t b !]
\centering
\includegraphics[width = 1 \columnwidth]{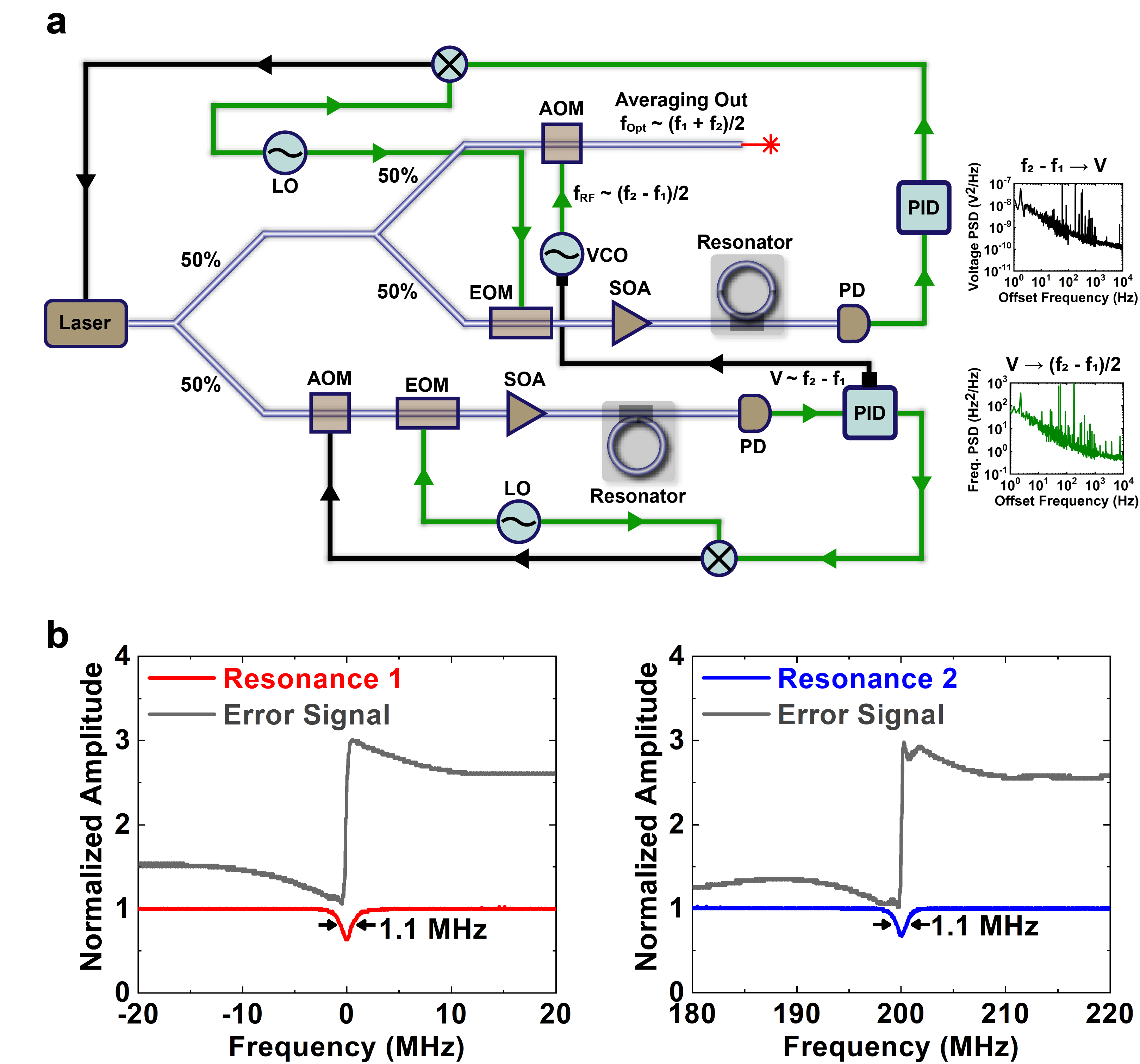}
\caption{
    \textbf{Optical Frequency Averaging System.}
    \textbf{a}, Diagram of the averaging system comprising acousto-optic modulators (AOM), electro-optic modulators (EOM), semiconductor optical amplifiers (SOA), photodiodes (PD), proportional integral derivative servo controllers (PID), and RF local oscillators (LO). A single laser is common to both paths, and the VCO and AOM combination enable the operation of averaging. The frequency and voltage PSDs illustrate the one-to-one correspondence between frequency and voltage.
    \textbf{b}, Scan over the individual resonances of the two resonators used in averaging. The PDH error signal is also shown.
}
\label{fig:fig2}
\end{figure}


\begin{figure}[t b !]
\centering
\includegraphics[width = 1 \columnwidth]{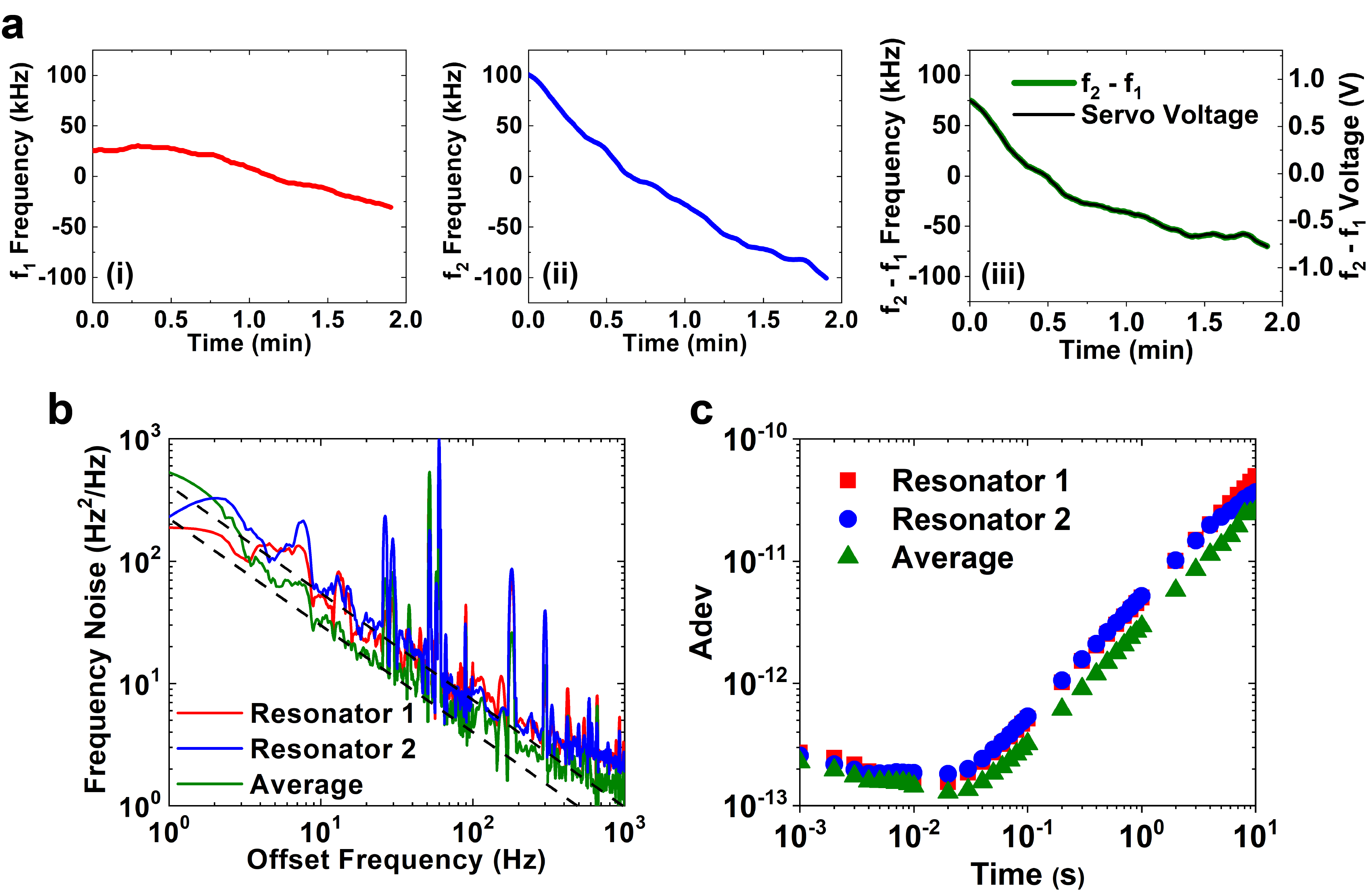}
\caption{
    \textbf{Measurement results.}
    \textbf{a}, Time series of the frequency excursion corresponding to each of the individual fiber-cavity stabilized lasers. In panel (iii), the calculated difference in frequency excursions is compared to the voltage derived from the servo output.
    \textbf{b}, Frequency noise measured for lasers stabilized to resonators 1 (red) and 2 (blue). The measured frequency noise of the average (green) is also provided. The dashed lines serve as guides for the falloff of noise with offset frequency.
    \textbf{c}, Allan deviation (Adev) of the cavity-stabilized lasers and of the averaged signal.
}
\label{fig:fig3}
\end{figure}

Figure 3a illustrates further the correspondence between optical frequency and voltage. Panels (i) and (ii) show the frequency excursions that each optical signal takes when stabilized to separate optical fiber cavities. Over a period of 2 minutes, each optical signal drifts independently on the scale of 100 kHz and also exhibits uncorrelated frequency fluctuations on faster time scales. Panel (iii) compares the calculated frequency difference between the two optical signals to the voltage experimentally derived from the AOM servo loop. The excellent agreement between voltage and frequency demonstrates our ability to convert from one domain to the other and back without losing information.

The frequency noise spectral density (Fig. 3b) highlights the noise reduction gained from averaging the frequency fluctuations across two fiber resonators. Each cavity-locked laser exhibits a characteristic decrease in noise as the offset frequency increases, as is indicated by the dashed line guides. The two individual optical signals have similar characteristics and noise levels, which makes them ideal for averaging. Should the noise be asymmetric between the two resonators at any offset frequency, the average will be heavily influenced towards the signal with higher noise. At 10 Hz offset, the individual cavity-locked lasers exhibit frequency noise at the level of 55 Hz\begin{math}^2\end{math}/Hz, while the averaged signal exhibits noise at the level of 29 Hz\begin{math}^2\end{math}/Hz. This reduction gained through averaging is close to the ideal factor of 2 expected for measurements of PSD.

The fractional frequency measurements (Adev) of Fig. 3c demonstrate a similar improvement to noise achieved by averaging. At 20 ms, each of the individual resonator-locked lasers exhibit a fractional fluctuation of \begin{math} 1.8\times10^{-13} \end{math}, which corresponds to a frequency variation of 40 Hz. The averaged output improves the Adev to \begin{math} 1.3\times10^{-13} \end{math} or 29 Hz frequency. These values of frequency fluctuation also agree closely with integrated linewidths of 50 Hz and 34 Hz derived from Fig. 3b for the cases without and with averaging applied.

Thus far, the sources of noise involved in averaging have been similar in magnitude but uncorrelated in phase between both optical signal paths. For arbitrary fluctuating noise sources \begin{math} n_1 \end{math} and \begin{math} n_2 \end{math} on optical signals 1 and 2, the noise propagates into the averaged signal through \begin{math} f_{avg} = ((f_1 + n_1) + (f_2 + n_2))/2 \end{math}. For the case of uncorrelated noise sources, \begin{math} n_1 \end{math} and \begin{math} n_2 \end{math} cannot be directly combined, and must be converted to power spectral density first before being summed together. This conversion squares the expression for \begin{math} f_{avg} \end{math}, and with uncorrelated cross-terms between signals 1 and 2 averaging to zero, the noise of the averaged spectrum is half that of each optical signal individually. Converting back to linear units, the frequency fluctuations on \begin{math} f_{avg} \end{math} are reduced by a factor of \begin{math} \sqrt{2} \end{math}. However, when the noise sources are completely correlated, they can instead be directly summed in linear units. After division by 2 in the expression for \begin{math} f_{avg} \end{math}, no effects of noise reduction would be observed after averaging. Another case worth considering is when the noise is only present on one optical signal. This noise would be directly reduced by a factor of 2 in the averaged signal, which is larger than the \begin{math} \sqrt{2} \end{math} reduction in the uncorrelated noise case.

We investigate the cases of asymmetric noise and correlated noise by experimentally introducing these noise types into our system in Fig. 2a. Figure 4a demonstrates the effects of averaging for the case of noise being present only on Resonator 1. This noise is created through an 800 Hz current modulation of the SOA in Resonator 1's signal path, which manifests as a frequency shift via a change of refractive index through thermal effects and the Kerr nonlinearity. The noise peak on Resonator 1 appears at a level of 61 $\text{Hz}/\sqrt{\text{Hz}}$, whereas a residual peak of 3 $\text{Hz}/\sqrt{\text{Hz}}$ appears on Resonator 2. This peak results because the noise modulation on the first path is written onto the seed laser, which is also sent into path 2. The second path's servo is unable to completely suppress the effects of this modulation. However, as this parasitic peak is ~20$\times$ lower than that of the primary path, it can be effectively ignored in the averaging. The averaged signal exhibits a peak of 32 $\text{Hz}/\sqrt{\text{Hz}}$, which is approximately half the peak of Resonator 1, as is expected.

Next we apply correlated noise to both resonators via a common audio tone received by both optical systems. For an applied tone at 800 Hz, Fig. 4b indicates a peak of 16 $\text{Hz}/\sqrt{\text{Hz}}$ for Resonator 1, whereas the peak of Resonator 2 appears at the level of 14 $\text{Hz}/\sqrt{\text{Hz}}$. In this case, the resulting peak on the averaged signal is 13 $\text{Hz}/\sqrt{\text{Hz}}$, or approximately equal to that of the base optical signals used in the averaging. Here, the direct correlation in noise on both optical signals results in no perceivable noise reduction through averaging.


\begin{figure}[t b !]
\centering
\includegraphics[width = 1 \columnwidth]{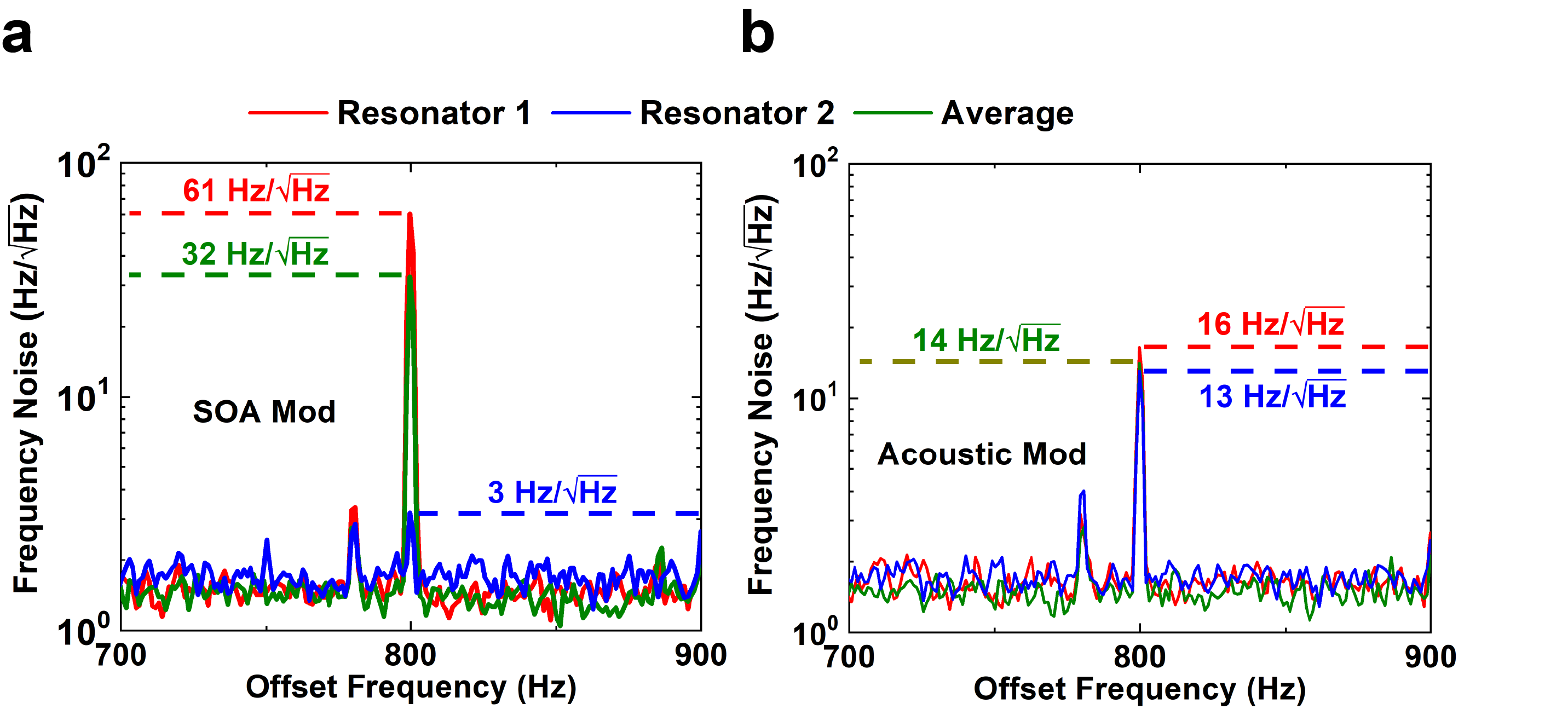}
\caption{
    \textbf{Correlation Effects on Noise Averaging.}
    \textbf{a}, Noise averaging when intentional SOA modulation is introduced on only one of the optical signals.
    \textbf{b}, Noise averaging when correlated acoustic noise is introduced on both optical signals.
}
\label{fig:fig4}
\end{figure}

\section{Conclusion}

The ability to apply statistical averaging to light is promising as a way to further narrow the linewidth of lasers beyond limits imposed by thermal noise or other fundamental noise sources. We demonstrated here two fiber cavity-stabilized lasers, each exhibiting fluctuations of 40 Hz, averaged together to produce an output having frequency fluctuations of 28 Hz. In contrast to techniques where the mode volume is simply increased, our method of averaging is applicable to arbitrary optical signals having any distribution of noise. This averaging scheme and experimental approach can also be readily extended to greater quantities of participating systems by sampling more systems in parallel and changing the mathematical operations to maintain self-consistency. In this way, multiple resonators may be stacked vertically on a single semiconductor chip and averaged together, to fully take advantage of all three dimensions as available space.


\bibliography{Optical_Averaging}

\end{document}